\documentclass{nature}

\usepackage[utf8]{inputenc}
\usepackage{graphicx}
\usepackage{aa_sym}

\usepackage{amssymb}
\usepackage{amsmath}
\usepackage{mathrsfs}
\usepackage{color}

\usepackage{lineno}

\newcommand{\lya}{Ly$\alpha$}
\newcommand{\hi}{H~{\small I}}
\newcommand{\mgi}{Mg~{\small I}}
\newcommand{\mgii}{Mg~{\small II}}
\newcommand{\mgiia}{\mgii~$\lambda$2796}
\newcommand{\mgiib}{\mgii~$\lambda$2803}
\newcommand{\mgiiab}{\mgii~$\lambda\lambda$2796,2803}
\newcommand{\civ}{C~{\small IV}}
\newcommand{\heii}{He~{\small II}}
\newcommand{\ciii}{C~{\small III}]}

\title{ Bipolar Outflows out to 10~kpc for Massive Galaxies at Redshift $z\approx 1$ \\
}

\author{Yucheng Guo$^{1}$, Roland Bacon$^{1}$, Nicolas F. Bouché$^{1}$, Lutz Wisotzki$^{2}$, Joop Schaye$^{3}$, Jérémy Blaizot$^{1}$, Anne Verhamme$^{4}$, Sebastiano Cantalupo$^{5}$, Leindert Boogaard$^{6}$, Jarle Brinchmann$^{7,3}$, Maxime Cherrey$^{1}$, Haruka Kusakabe$^{4}$, Ivanna Langan$^{1,8}$, Floriane Leclercq$^{9}$, Jorryt Matthee$^{10}$, L\'eo Michel-Dansac$^{1}$, Ilane Schroetter$^{11}$, Martin Wendt$^{2,12}$}

\begin{document}
\setlength{\unitlength}{1mm}
\sloppy

\maketitle
{\footnotesize
\begin{affiliations}
\item{Univ Lyon, Univ Lyon1, Ens de Lyon, CNRS, Centre de Recherche Astrophysique de Lyon UMR5574, F-69230, Saint-Genis-Laval, France}
\item{Leibniz-Institut f{\"u}r Astrophysik Potsdam (AIP), An der Sternwarte 16, 14482 Potsdam, Germany}
\item{Leiden Observatory, Leiden University, P.O. Box 9513, 2300 RA Leiden, The Netherlands}
\item{Observatoire de Gen\`eve, Universit\'e de Gen\`eve, 51 Ch. des Maillettes, CH-1290 Versoix, Switzerland }
\item{Dipartimento di Fisica ``G. Occhialini'', Universit\`a degli Studi di Milano Bicocca, Piazza della Scienza 3, 20126 Milano, Italy}
\item{Max Planck Institute for Astronomy, K\"onigstuhl 17, 69117, Heidelberg, Germany}
\item{Instituto de Astrof{\'\i}sica e Ci{\^e}ncias do Espaço, Universidade do Porto, CAUP, Rua das Estrelas, PT4150-762 Porto, Portugal}
\item{European Southern Observatory, Karl-Schwarzschild-Str. 2, D-85748, Garching, Germany}
\item{Department of Astronomy, University of Texas at Austin, 2515 Speedway, Austin, TX 78712, USA}
\item{Department of Physics, ETH Zürich, Wolfgang-Pauli-Strasse 27, 8093 Zürich, Switzerland}
\item{GEPI, Observatoire de Paris, PSL Universit\'e, CNRS, 5 Place Jules Janssen, 92190 Meudon, France}
\item{Institut f{\"u}r Physik und Astronomie, Universit{\"a}t Potsdam,Karl-Liebknecht-Str. 24/25, D-14476 Golm, Germany}
\end{affiliations}

}
\begin{abstract}
Galactic outflows are believed to play a critical role in the evolution of galaxies by regulating their mass build-up and star formation \cite{tumlinson17}.
Theoretical models assumes bipolar shapes for the outflows that extends well into the circumgalctic medium (CGM), up to tens of kpc perpendicular to the galaxies. They have been directly observed in the local Universe in several individual galaxies, e.g., around the Milky Way and M82 \cite{predehl20,bland88}.
At higher redshifts, cosmological simulations of galaxy formation predict an increase in the frequency and efficiency of galactic outflows due to the increasing star formation activity \cite{muratov15}.
Outflows are responsible for removing potential fuel for star formation from the galaxy, while at the same enriching the CGM and the intergalactic medium.
These feedback processes, although incorporated as key elements of cosmological simulations, are still poorly constrained on CGM scales. 
Here we present an ultra-deep MUSE image of the mean \mgii\ emission surrounding a sample of galaxies at $z \approx 1$ that strongly suggests the presence of outflowing gas on physical scales of more than 10~kpc. We find a strong dependence of the detected signal on the inclination of the central galaxy, with edge-on galaxies clearly showing enhanced \mgii\ emission along the minor axis, while face-on galaxies display much weaker and more isotropic emission. We interpret these findings as supporting the idea that outflows typically have a bipolar cone geometry perpendicular to the galactic disk.
We demonstrate that the signal is not dominated by a few outliers.
After dividing the galaxy sample in subsamples by mass, the bipolar emission is only detected in galaxies with stellar mass $\mathrm{M_* \gtrsim 10^{9.5} M_\odot}$.

\end{abstract}
\clearpage

\clearpage
During the past ten years, integral field spectrograph (IFS) facilities such as VLT/MUSE \cite{bacon10} and Keck/KCWI \cite{morrissey18} have enabled mapping the CGM or even the IGM in emission using several tracers such as Ly$\alpha$ \cite{wisotzki16, wisotzki18, leclercq17, cai19, bacon21, kusakabe22}, \civ, \heii\ and \ciii\ \cite{guo20}, and other rest-frame optical emission lines \cite{johnson22, herenz22} out to very large distances from galaxies.
Yet \lya\ emission can only be observed by ground-based facilities at $z\gtrsim 2$, and due to the severe effects of cosmological surface brightness dimming it is very difficult to obtain spatially resolved spectroscopy at such high redshifts.

The \mgiiab\ doublet presents a potential alternative to \lya\ for tracing the CGM.
Due to its resonant nature, extended \mgii\ emission is expected to occur under similar conditions as \lya.
The ionization potential of \mgi\ (7.6~eV) is lower than that of \hi\ (13.6~eV), implying that the CGM remains at least partly neutral when \mgii\ is produced.
This makes \mgii\ a more promising tracer of the cool and metal-enriched CGM gas compared to \lya.
The line can be observed from the ground at redshifts $z\gtrsim 0.1$ \cite{kacprzak13}, facilitating higher linear resolution and much lower surface brightness dimming.
Nevertheless, the detection of circumgalactic \mgii\ is challenging because \mgii\ is intrinsically much fainter than \lya. 
So far, extended \mgii\ emission was detected around two galaxies by long slit spectra \cite{rubin11,martin13}, and also more recently by IFS \cite{burchett21,zabl21,leclercq22}.
The occurrence frequency and outflow nature of the extended \mgii\ emission are still unclear, and therefore we do not know if it is a common feature of galaxies across the mass spectrum.

We answer this question using extremely deep MUSE observations of a sample of galaxies at $z \approx 1$ in the Hubble Ultra Deep Field \cite{bacon23}. We employ HST images to construct a sample of 112 edge-on and 60 face-on galaxies. 
We perform an oriented stacking of the MUSE datacube segments after aligning the each edge-on galaxy along the direction of the photometric major axis.
We stack the face-on galaxies by their original sky orientations. 
Before stacking, we remove the continuum and mask bright neighboring objects in each individual datacube (Methods).

\begin{figure}
\begin{center}
\includegraphics[width=0.65\textwidth]{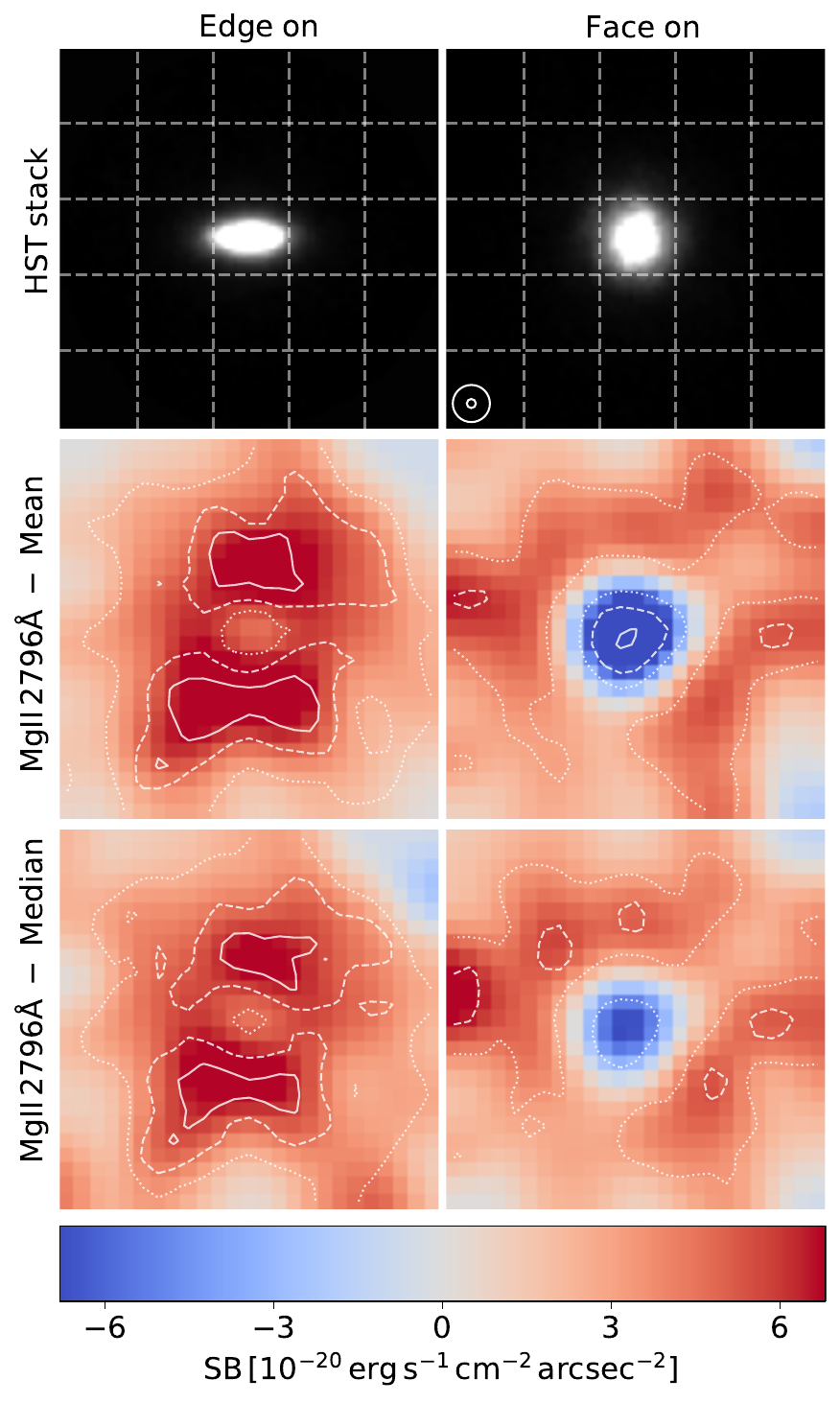}
\end{center}
\caption[]{\textbf{Maps of the stellar continuum and \mgii\ emission for our galaxy sample, with edge-on and face-on galaxies shown in the left and right columns, respectively.} 
The top row displays the stacked average of HST images (F435W, F775W, F606W, F850LP and F160W bands), representing the stellar component.
The middle and bottom rows show, respectively, the mean and median stacked \mgii\ 2796 \AA\ pseudo-NB images, smoothed by a Gaussian kernel of width of $0.4''$.
The white contours correspond to \mgii\ significance levels of 2, 4 and 6 $\sigma$ (dotted, dashed and solid, respectively).
Each thumbnail has a size of $5''\times5''$, corresponding to about $40\times 40$~kpc at the median redshift of the sample.
Each grid cell at the top row corresponds to a distance of about 8~kpc.
The smaller and larger circles in the upper right panel represent the PSF FWHM of HST and MUSE, respectively.
We extract spectra from each grid cell shown in the top panels, and present them in the corresponding panels in Extended Data Fig.~\ref{fig_stack_spec}.   \label{fig_stack_img}} 
\end{figure}

We then construct \mgiia\ pseudo-narrowband (NB) images by summing the stacked datacubes in the spectral direction, followed by spatial filtering. 
Fig.~\ref{fig_stack_img} shows the stacking results of the HST images (stellar component) and of the \mgiia\ line emission. We clearly detect resolved \mgii\ emission extending over CGM scales.
The scale of each thumbnail is $5\arcsec  \times 5\arcsec$ (corresponding to $\approx$ 40 $\times$ 40~kpc at the median redshift of the sample). 

The stack of edge-on galaxies in the left column of Fig.~\ref{fig_stack_img} shows a clear enhancement of \mgii\ emission along the minor axis.
We detect a bipolar shape extending out to a radius of about 10~kpc in the stacked cube, suggesting that this is due to gas recently ejected from the galaxy.
This anisotropic emission is seen in both the mean and median stacked cubes, indicating the \mgii\ outflows are common among our galaxy sample.
We verify that this anisotropic pattern does not result from a few bright outliers, but is a generic property of galaxies in our sample (Methods).

We do not detect strong extended emission around the face-on galaxies. 
There is however strong \mgii\ absorption in the central region coinciding with the stellar component of the galaxies.
In the outer region we see a weak ring pattern in both the mean and median stacks. 
The detection of the ring is robust (Methods).
The physical origin for this pattern is unclear.
It might be attributed to an isotropic \mgii\ halo, to inflowing or re-accreted gas, or to a face-on outflow cone extending to even larger radii.
Further observations and simulations are needed for a better understanding. 

\begin{figure}
\begin{center}
\includegraphics[width=0.7\textwidth]{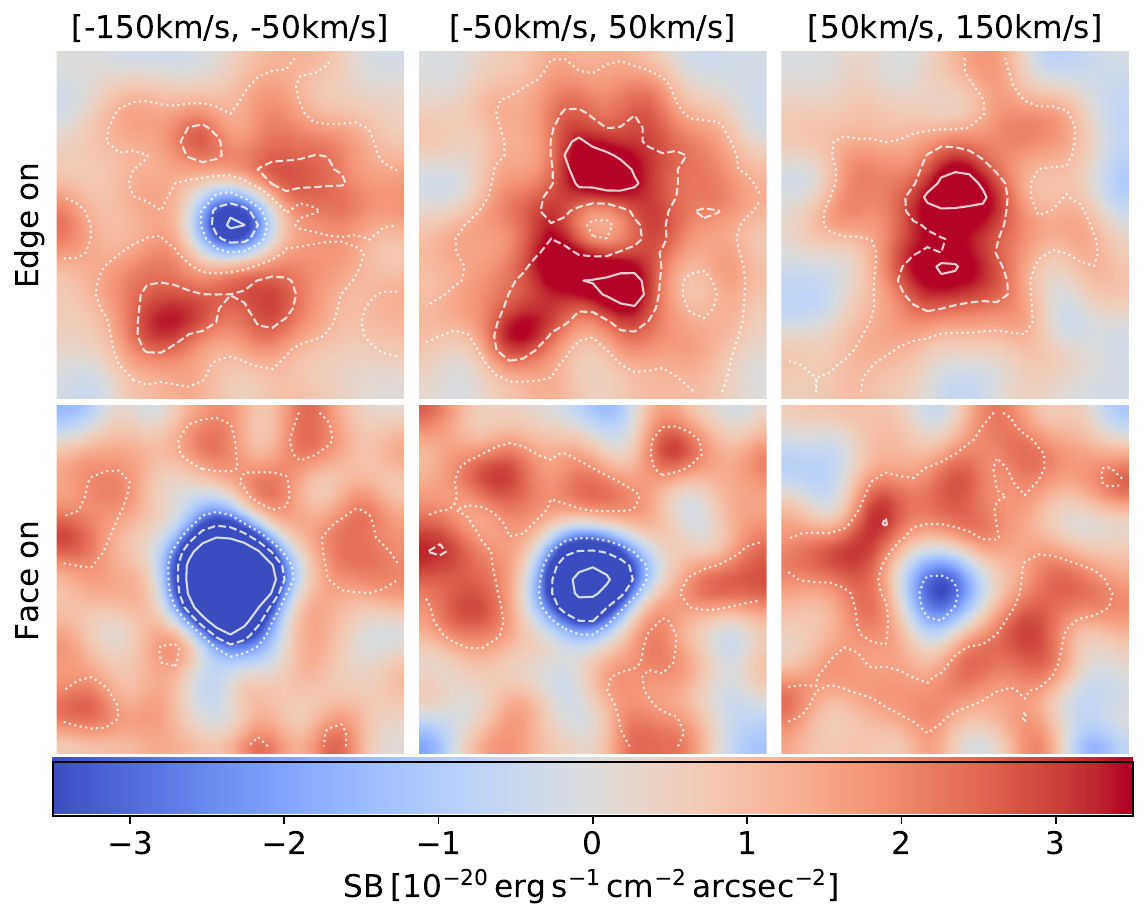}
\end{center}
\caption[]{\textbf{Pseudo-NB images for different velocity bins.}
The images are extracted in intervals of 100~km/s ($\approx$1\AA\ in the rest frame) centered at \mgii\ 2796\AA, then smoothed by a Gaussian of width of 0.4''. The upper and lower rows show the edge-on and face-on samples, respectively.
The white contours correspond to \mgii\ significance levels of 2, 4 and 6 $\sigma$ (dotted, dashed and solid, respectively).
All scales and annotations are as in Fig.~\ref{fig_stack_img}.
\label{fig_vbins}} 
\end{figure}

To visualize the kinematic structure of the extended \mgii\ emission, we split the wavelength range of the NB image of Fig.~\ref{fig_stack_img} into three velocity intervals of 100~km/s, shown in Fig.~\ref{fig_vbins}.
For edge-on galaxies, the anisotropic \mgii\ emission is seen in all velocity intervals, although it is less evident in the bluest channel.
At the velocity centered on the systemic velocity we detect strong emission both above and below the galaxy disk.
Towards positive velocities, where there is no central absorption any more, \mgii\ is still strong along the minor axis.
This shows that the anisotropic \mgii\ emission is an inherent property of the CGM and independent of the central absorption of the stellar component of the galaxy.
The face-on sample shows central absorption at all velocity intervals. It is strongest in the most negative velocity range and gets successively weaker towards redder wavelengths.
At all velocities the central absorption is stronger for face-on galaxies than for edge-on systems.

\begin{figure*}[ht!]
\includegraphics[width=0.496\textwidth]{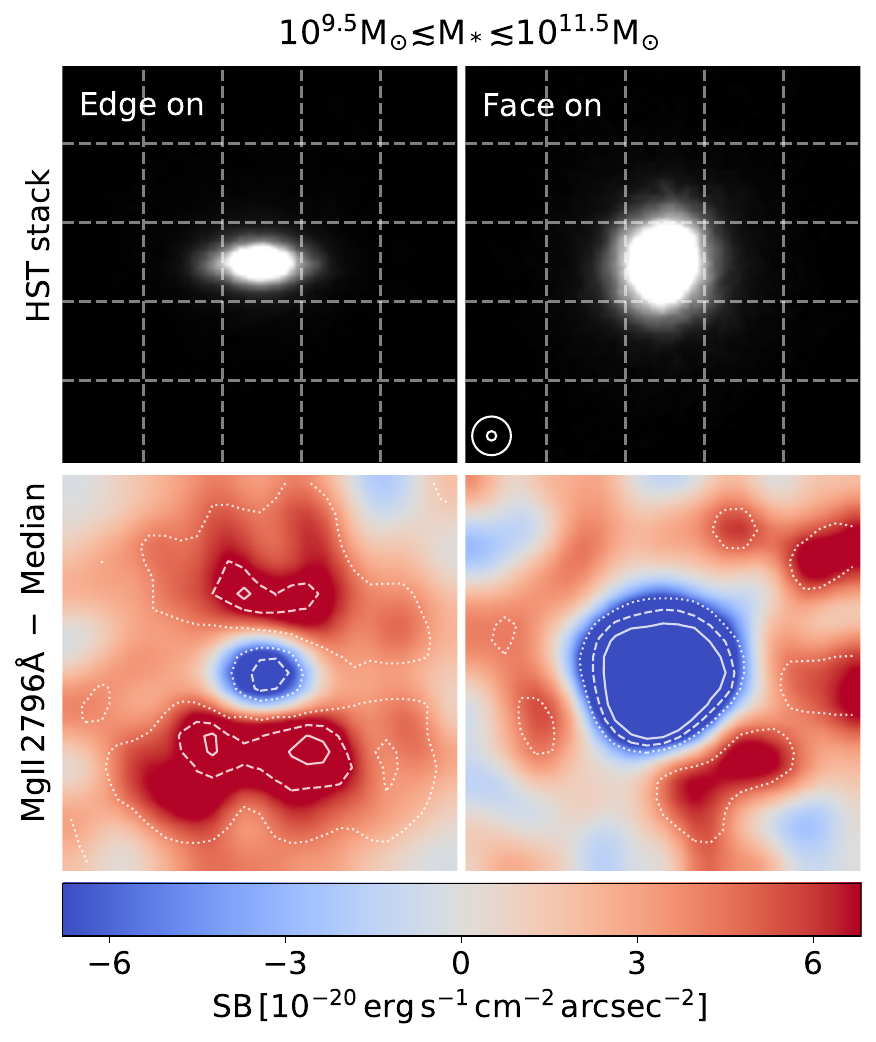}
\includegraphics[width=0.495\textwidth]{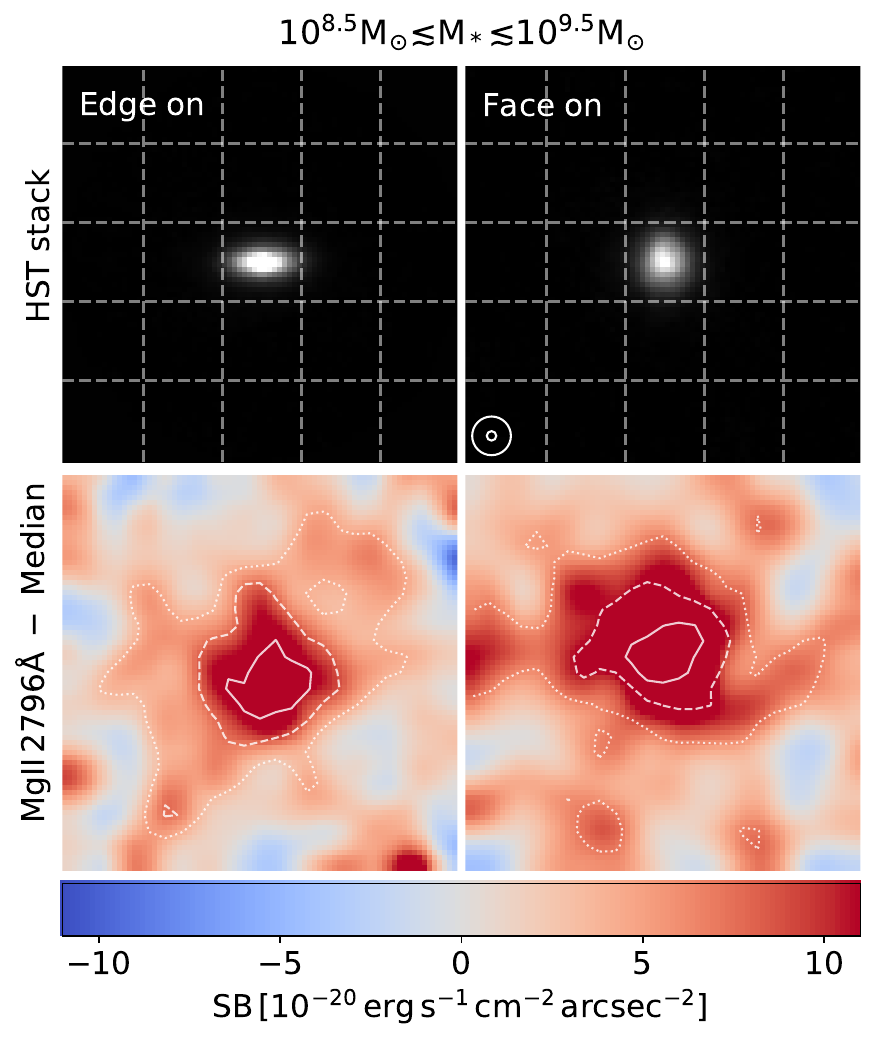}
\caption[]{\label{fig_mass_bins}
\textbf{Maps of the stellar continuum and median stacked \mgii\ emission for the high-mass (left) and the low-mass (right) subsamples}, bisected at the median stellar mass of $\mathrm{10^{9.5} \, M_\odot }$.
The bottom row shows the  \mgii\ 2796 \AA\ pseudo-NB images. 
The color scales of the left and right subplots were adjusted for presentation purposes.
The white contours correspond to \mgii\ significance levels of 2, 4 and 6 $\sigma$ (dotted, dashed and solid, respectively).
All scales and annotations are as in Fig.~\ref{fig_stack_img}.
}
\end{figure*}

In order to study the stellar mass dependence of the \mgii\ emission, we split our sample into halves around the median mass of the sample $\mathrm{10^{9.5} \, M_\odot}$. 
As shown in Fig.~\ref{fig_mass_bins}, only the high-mass edge-on subsample displays the anisotropic \mgii\ emission. 
The corresponding high-mass face-on subset, in contrast, shows very strong central absorption surrounded by a noisy emission ring-like feature.
On the other hand, the low-mass subsample does not display such anisotropic emission, with both the edge-on and the face-on orientations presenting emission of irregular morphology. 
This could hint a more irregular geometry of outflows in the low-mass systems.
However, as a caveat we note that low-mass galaxies are smaller, implying that the impact of the PSF is more pronounced so that smaller-scale wind signatures would be more difficult to detect. 
A correlation between galaxy \mgii\ equivalent width and stellar mass was previously reported \cite{kornei12,feltre18}, that the \mgii\ emitters generally have lower masses than \mgii\ absorbers, with a transition occurring around $\mathrm{10^{9.5} \, M_\odot}$.

\begin{figure}
\begin{center}
\includegraphics[width=0.7\textwidth]{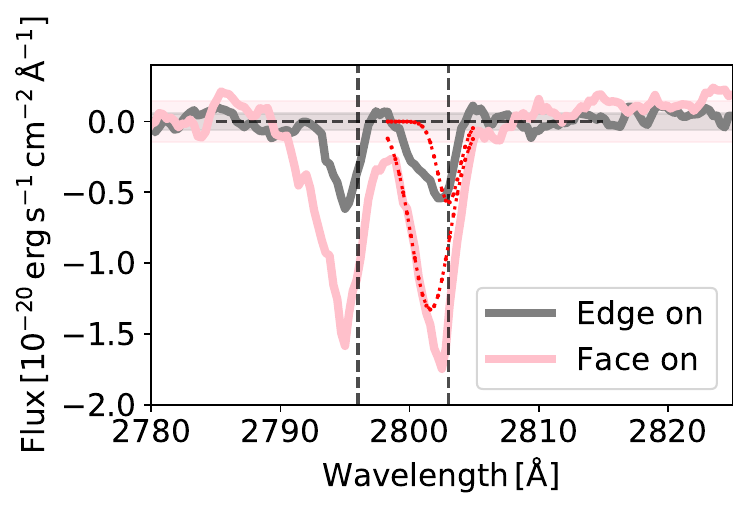}
\end{center}
\caption[]{\textbf{Continuum-subtracted mean spectra of the high-mass galaxies.}
The spectra are extracted from the central 1''.
The pink and grey spectra show the face-on and edge-on galaxy subsamples, respectively.
The dotted red lines present the double Gaussian decomposition of the \mgii\ 2803\AA\ line (Methods).
The two vertical dashed lines denote the wavelengths of the \mgii\ doublet. The horizontal dashed line denotes the zero flux level. 
The colored shadings represent the 1$\sigma$ error ranges of the corresponding spectra.
The equivalent widths (EWs) of the \mgii\ 2796 \AA\ line for the face-on and edge-on galaxy are $\mathrm{ 7.4 \pm 0.9 \, \AA }$ and $\mathrm{ 2.5 \pm 0.7 \, \AA }$, respectively.
\label{fig_spec_outflow2}} 
\end{figure}

In Fig.~\ref{fig_spec_outflow2} we compare the spectra extracted from the central regions of the high-mass galaxies.
Evidently the \mgii\ absorption profile of the face-on galaxies is much broader than that of the edge-on galaxies. 
If we assume that the two samples of galaxies have similar properties except for their inclination, then the widening of the absorption line indicates the typical velocity of the outflow ($v_{\mathrm{out}}$).
We use two independent methods to measure this velocity, obtaining a consistent estimate of $v_{\mathrm{out}}\mathrm{\approx 180 \, km/s}$ (Methods).

Galactic outflows are usually of low gas density and low surface brightness, and therefore difficult to observe in emission, especially towards high redshifts.
In previous studies, evidence for the galactic outflows and their connection to the CGM at $z\gtrsim 1$ is mainly provided indirectly by absorption diagnostics against bright background sources. For example, recent observations find that \mgiiab\ absorption is not evenly distributed around galaxies, but occurs preferentially close to the major or minor axes of the galaxies \cite{bouche12,zabl19}. It is particularly strong along the minor axis, where the metal-enriched outflows are expected \cite{martin19,schroetter19}. However, direct spatially-resolved spectroscopic observations on such outflows as well as the frequency of their occurrence are largely missing. 
In this study, using ultra deep MUSE IFS observations in the Hubble Ultra Deep Field, we discover that the anisotropic \mgii\ emission extending up to $\approx$10~kpc is common for massive ($\mathrm{M_* \gtrsim 10^{9.5} M_\odot}$) edge-on galaxies at $z \approx 1$. 
This finding provides the most direct evidence of the prevalence of cool and metal-enriched galactic outflows at high redshift. 
Simulations predict that most of the metals produced inside high-redshift galaxies are carried out by galactic winds \cite{muratov17}. 
The galactic winds may produce an anisotropic metallicity distribution in the CGM \cite{peroux20}.
The results of this paper provide further support for the key role of galactic winds in the transport of metals from the galaxy to the CGM.


\begin{addendum}
 \item 
This is a pre-print of an article published in Nature. The final authenticated version is available online at: https://www.nature.com/articles/s41586-023-06718-w .

Y.G. thanks Ziyan Xu for helpful discussions.
Y.G., R.B. and L.W. acknowledge support from the ANR/DFG grant L-INTENSE (ANR-20-CE92-0015, DFG Wi 1369/31-1).
L.W. acknowledges support by the ERC Advanced Grant SPECMAG-CGM (GA101020943).
A.V. and H.K. acknowledge support from the Swiss National Fundation grant PP00P2\underline{\hspace{0.5em}}211023.
S.C. gratefully acknowledges support from the European Research Council (ERC) under the European Union’s Horizon 2020 Research and Innovation programme grant agreement No 864361.
L.B. acknowledges support by ERC AdG grant 740246 (Cosmic-Gas).
J. Brinchmann acknowledges financial support from the Fundação para a Ciência e a
Tecnologia (FCT) through research grants UIDB/04434/2020 and UIDP/04434/2020, national funds PTDC/FIS-AST/4862/2020 and work contract 2020.03379.CEECIND.
N.F.B. acknowledges support from the ANR grant 3DGasFlows (ANR-17-CE31-0017).

\item[Author contributions]
Y.G. conceived the project. 
R.B. led the MUSE data acquisition and data reduction. 
All authors participated in either the observation and data reduction of the MUSE Hubble Ultra Deep Field surveys.
Y.G. performed the sample selection and analysed
the data. 
Y.G., R.B., N.F.B., L.W., J.S., J.Blaizot, S.C. worked on the interpretation of the results.
Y.G. wrote the manuscript and produced the figures, with R.B., N.F.B., L.W., J.Blaizot contributing to their design.
All coauthors provided critical feedback to the text and helped shape the manuscript.

\item[Author information]
The authors declare no competing financial interests. Readers are welcome to comment on the online version of
the paper. Correspondence and requests for materials should be addressed to Y.G. (yucheng.guo@univ-lyon1.fr).

\item[Data availability]
This work is mainly based on DR2 of the MUSE Hubble Ultra Deep Field surveys. 
The reduced MUSE datacubes are available in Ref.~\cite{bacon23}.

\end{addendum}

\clearpage


\clearpage

\renewcommand{\figurename}{\textbf{Fig.}}

\clearpage

\clearpage

\clearpage

\clearpage


\begin{methods}

\section{Data reduction and analysis.}

This work is mainly based on data release 2 (DR2) of the MUSE Hubble Ultra Deep Field surveys \cite{bacon23}.
The DR2 data consists of 3 datasets, a $\mathrm{3 \times 3}$ arcmin$^2$ mosaic of 9 MUSE fields at 10-hour depth (hereafter MOSAIC), a $\mathrm{1 \times 1}$ arcmin$^2$ field at 31-hour depth (hereafter UDF-10), and the MUSE eXtremely Deep Field (MXDF), with a deepest achieved exposure of 141 hours.
This is the deepest spectroscopic survey ever performed, reaching an unresolved emission line $\mathrm{ 1\sigma }$ surface brightness limit of $\mathrm{ <10^{-19}\,erg\,s^{-1} \, cm^{-2} \, arcsec^{-2} } $.
These deep MUSE observations enable studies of extremely low surface brightnesses, such as spatially and kinematically resolved analysis of the CGM by \lya\ \cite{claeyssens19,leclercq20,erb22},  the detection of a cosmic web filament in \lya\ emission on scales of several cMpc \cite{bacon21}.

The DR2 catalogue provides redshifts, multi-band photometry, morphological and spectral properties, as well as measurements of stellar mass and star formation rate of all the galaxies discovered in the MOSAIC, UDF-10 and MXDF fields \cite{bacon23,bacon17}.
In this work, we focus on the \mgii\ doublet.
Given the wavelength range of MUSE, \mgii\ can be detected in the redshift interval of $\approx 0.70 - 2.30$.
There are 568 galaxies detected in this redshift range.

Since the goal of this work is to determine the azimuthal dependence of extended \mgii\ emission, a priori determination of galaxy orientation is required.
We selected subsamples of edge-on and face-on galaxies by visual inspection.
By cross-matching our visual selection with the GalPak measurement \cite{bouche21} that covers a subset of this galaxy sample, we find that the majority of the face-on and edge-on subsamples have inclination angles of around $<30^\circ$ and $>55^\circ$, respectively.
Even if this morphological classification is approximate, it is precise enough for our statistical analysis.
To remove potentially merging objects, we exclude all pairs of galaxies with small projected angular separations (2'') and small line-of-sight relative radial velocities ($\leqslant$1000~km/s).
We selected 112 edge-on and 60 face-on galaxies, with accumulated exposure times of 2599h and 1331h, respectively.

The redshift and stellar mass ($\mathrm{M_*}$) distributions of the parent sample and the edge-on and face-on subsamples are shown in Extended Data Fig.~\ref{fig_distribution}. 
The two subsamples are comparable to each other in redshift and mass.
The median redshift is $z \approx 1.1$. 
The median stellar mass is $\mathrm{ \approx 10^{9.5} M_\odot }$, with a $\mathrm{\lesssim }$0.1~dex difference between the edge-on and face-on subsamples.
We also note that this morphological classification is mainly reliable for galaxies with mass of $\mathrm{ M_* > 10^{8.5} M_\odot } $, 
because lower-mass galaxies are not well resolved,
even with ultra-deep HST images.
The HST images of galaxies in the face-on and edge-on subsamples are presented in Extended Data Fig.~\ref{fig_hst_fo} and ~\ref{fig_hst_eo}.
For the edge-on galaxies, we find the major axis for each target from the HST images via a principal component analysis.
In Extended Data Fig.~\ref{fig_hst_eo}, we present the measurements of the galaxy major and minor axes.

Despite the achieved depth of the sample, extended \mgii\ emission is detected in only a small fraction of galaxies \cite{zabl21}.
To increase the S/N, determine the average line strengths and obtain spatially resolved information on the properties of the CGM, we adopt a full 3-dimensional stacking procedure.
We stack the $5'' \times 5''$ MUSE mini-datacubes centered on each source.
Before stacking we remove the continuum by performing a spectral median filtering using a wide spectral window of 200~\AA.
This approach provides a fast and efficient way to remove continuum sources in the search for extended line emission.
We also mask bright neighboring objects to avoid possible contamination.
We shifted individual datacubes and re-binned them to a common (rest-frame) wavelength frame. 
We did not re-scale the flux of each individual datacube to correct for the impact of cosmological dimming.
The cubes are averaged and weighted by square root of their exposure time. 
Finally, we extract pseudo-NB images from the stacked cube.

In this work, we investigate the relation between galaxy orientation and extended \mgii\ emission. Therefore, for the edge-on galaxies, we stack the datacubes by re-aligning each one along the direction of the galaxy major axis. 
We then proceed with the same stacking procedure as described above using the re-aligned cubes.
For face-on galaxies, we stack the datacubes using their original orientations.

We apply a pseudo-NB filter centered on the \mgii\ 2796~\AA\ line, with a rest-frame full width of 3~\AA, corresponding to a velocity range of $\approx 300$~km/s.
In Fig.~\ref{fig_stack_img}, we show the first line (2796~\AA\ ) of the \mgii\ doublet. Anisotropic pattern can be clearly observed around the edge-on galaxy. 
To quantify the statistical significance of the observed pattern, we calculate the S/N as follows.
We estimate the noise by generating pseudo-NB images at 50 random wavelengths adjacent to the \mgii\ 2796~\AA\ line, within $\pm$80~\AA, while avoiding the wavelength range of the \mgii\ doublets.
The widths of these random NBs are the same as the NB for \mgii\ line.
Using these random pseudo-NB images, we determine the noise and calculate the corresponding S/N. 
The S/N of the \mgii\ detection is indicated by the contours in Fig.~\ref{fig_stack_img}.

Along the minor axis of the edge-on galaxy, the anisotropic \mgii\ emission reaches a peak S/N of 6$\sigma$.
The S/N for the face-on sample is lower, but we will still show the robustness of the ring pattern in Extended Data Fig.~\ref{fig_spec_outflow1}.

We also plot the \mgii\ 2796~\AA\ maps at narrower velocity bins that correspond to line-of-sight velocities of $\Delta v =$ $-$150 to $-$50, $-$50 to 50, and 50 to 150~$\mathrm{ km/s }$ (Fig.~\ref{fig_vbins}).
The sum of the three velocity bins corresponds to approximately 3~\AA\ in rest frame, so this combination of the three maps equals the pseudo NB in Fig.~\ref{fig_stack_img}.
In Fig.~\ref{fig_vbins} and Fig.~\ref{fig_mass_bins}, we calculate the S/N contours following the same method as mentioned above.

\section{The size of the anisotropic \mgii\ emission.}

To further quantify the spatial scale of the anisotropic \mgii\ emission around edge-on galaxies, we measure the distance ($\mathrm{b}$) of the bipolar \mgii\ emission to the galaxy disk.
This is achieved by calculating the flux-weighted distance to the galaxy disk for the pixels within the 6~$\sigma$ contours in Fig.~\ref{fig_stack_img}. 
The distance derived from the mean-stacked datacube is $\mathrm{ b = 9.6 \pm 1.7 \, kpc }$, while the median-stacked datacube yields a distance of $\mathrm{ b = 9.4 \pm 2.0 \, kpc }$.
To determine the errors associated with these distances, we replicate the procedure after adding noise to produce 100 mock NB images and repeat the distance calculation. 

For comparison, we also compute the size of the stellar emission region. We perform the stacking procedure on the MUSE white light images of the edge-on galaxies, applying the same smoothing kernel as before. Subsequently, we fit a Sersic profile to the stacked image.
The resulting minor-axis effective radius is $\mathrm{ 3.2 \pm 0.3 \, kpc }$.

Additionally, we calculate the size of the anisotropic \mgii\ emission for the high-mass subsample in Fig.~\ref{fig_mass_bins} using the same methodology.
In this case, we utilize the 4~$\sigma$ contours to define the emission region.
The distance of the bipolar \mgii\ emission to the galaxy disk is $\mathrm{ b = 10.2 \pm 2.5 \, kpc }$.
Furthermore, we measure the minor-axis effective radius of the stellar continuum to be $\mathrm{3.6 \pm 0.3 \, kpc}$.

Another important morphological parameter characterizing galactic outflows is the opening angle ($\theta$). 
Despite the fact that galactic outflows in the local Universe do not commonly exhibit perfectly conical shapes in real observations, approximating them with an opening angle can simplify the modeling process.
In this work, we are able to directly measure this parameter from the edge-on image, thanks to the pre-selection based on galaxy inclination. 
In Fig.~\ref{fig_stack_img}, we define the extent of the outflow using a flux level that encompasses half of the peak flux ($\mathrm{\approx 5 \times 10^{-20} \, erg \, s^{-1} \, cm^{-2} arcsec^{-2} }$).
For the mean and median stacks, we measure average outflow opening angles of $\mathrm{ \theta \approx 68 \pm 8^{\circ } }$ and $\mathrm{ \theta \approx 70 \pm 11^{\circ } }$, respectively.
Similarly, employing the same method, we calculate the outflow opening angle for the high-mass subsample shown in Figure~\ref{fig_mass_bins} to be $\mathrm{\theta \approx 78 \pm 15^{\circ}}$.
It is important to note that the value of $\theta$ depends on the PSF and the smoothing kernel used, meaning that the actual $\theta$ is likely smaller than the measured value.
Nonetheless, it is worth mentioning that our measurement of $\theta$ is consistent with previous absorption studies \cite{schroetter19,rubin14} and is also close to the opening angle measured for the nearby starburst galaxy M82 \cite{walter02}.

\section{The prevalence of anisotropic \mgii\ emission.}
We have seen the anisotropic \mgii\ emission in both the mean and median stacks (Fig.~\ref{fig_stack_img}), indicating that the \mgii\ outflows are common among the galaxies in our sample.

To further demonstrate that the stacking is not dominated by a few outliers, we perform the following analysis.
We produce pseudo-NB images for each individual galaxy, using the same NB width as described above. 
Then we perform aperture photometry on each pseudo-NB image, placing the apertures 1'' above and below the galaxy plane along the minor axis. 
In this way we measure the fluxes of the regions where the bipolar outflows are expected.
We measure the S/N of the photometry. 
The distribution of the surface brightness and S/N in the outflow region for the edge-on galaxy sample is shown in Extended Data Fig.~\ref{fig_distribution_outflow}.
The smooth shape of the probability distribution indicates that the signal is not dominated by outliers.
The median surface brightness of the \mgiia\ outflow emission is $\mathrm{ 7.9 \times 10^{-20}\,erg\,s^{-1} \, cm^{-2} \, arcsec^{-2} } $.
We have also tried to remove the top 5\% of the edge-on galaxy sample with the highest S/N.
Stacking the remaining datacubes then results in a bipolar pattern very similar to Fig.~\ref{fig_stack_img}.

\section{The spectroscopic properties of the extended \mgii\ emission.}
We extract the spectra from different regions in the field of view.
The spectra are shown in Extended Data Fig.~\ref{fig_stack_spec}, with each panel corresponding to a grid cell at the top row in Fig.~\ref{fig_stack_img}.
The stack represents a total exposure time of thousands of hours, achieving a 1$\sigma$ noise level of less than $\mathrm{10^{-21} \, erg \, s^{-1} \, cm^{-2} \AA^{-1} }$.
We see prominent signals particularly in grid cells No.~8, 13, 18. 
For edge-on galaxies, in the regions along the galaxy minor axis (e.g., grid cells No.~8 and 18), the stack shows pure emission near the systemic redshift, supporting a scenario of galactic outflow along the minor axis.
The \mgii\ emission line doublets show a variation of line ratio over the field of view, indicating a complicated radiative transfer process.

The center region (e.g., cell No.~13) of the edge-on sample shows a complex combination of blueshifted absorption and redshifted emission. 
Previous works found that \mgii\ spectra of galaxies appear to show pure absorption for high-mass galaxies, emission for lower-mass galaxies, and P-Cygni like profiles for intermediate-mass galaxies \cite{feltre18}. 

\section{The ``ring'' pattern in the face-on galaxies.}

In Fig.~\ref{fig_stack_img} and Extended Data Fig.~\ref{fig_stack_spec}, we detect pure absorption in the center of the stack of face-on galaxies.
In the outer region, we see a weak ``ring" pattern in the NB images. 

In order to quantify the robustness of the ``ring" pattern, we show the spectra extracted from the stacked face-on sample (red line, extracted from an annular aperture of 1'' - 2'') in Extended Data Fig.~\ref{fig_spec_outflow1}. 
Although the absorption is strong, the redshifted \mgiia\ line is evident, with S/N $\approx$3.2. 
The EW of the \mgii\ 2796 \AA\ line is $\mathrm{ -21 \pm 15 \, \AA }$.
The physical origin for this pattern we see in the stack of face-on galaxies is unclear.
Due to the uncertainty in measuring the inclination angle, the face-on sample is likely not purely face-on. 
The ``ring" could also be attributed to inflowing or re-accreting gas, or outflows extending to large radii.
Further observations and simulations are needed for a better understanding. 

For comparison, in Extended Data Fig.~\ref{fig_spec_outflow1} we also plot spectra extracted from the bipolar outflow region (1.5'' upper and lower than the galaxy disk, with an aperture size of 1''). 
The EW of the \mgii\ 2796 \AA\ line is $\mathrm{ -56 \pm 22 \, \AA }$. 
Both lines of the \mgii\ doublet are obvious.
The line ratio of the doublet is $\mathrm{ 2.3 \pm 0.5 }$, which is compatible with the value of 2 expected for an optically thin gas.
The variation of the line ratio for extended \mgii\ emission has been reported \cite{chisholm20}.
Though with low S/N, we see variation of the \mgii\ doublet ratio (Extended Data Fig.~\ref{fig_stack_spec}) that may result from the complicated radiative transfer of the \mgii\ photons, and possibly hints the variation of gas density in an optically-thin scenario \cite{chisholm20,katz22}.

\section{The outflow in down-the-barrel absorption.}
It is also interesting to compare the down-the-barrel spectra from the galaxies themselves.
We compare the spectra extracted from the center of the high-mass galaxy sample in Fig.~\ref{fig_spec_outflow2}.
The EW of the \mgii\ 2796 \AA\ line of the face-on galaxy is $\mathrm{ 7.4 \pm 0.9 \, \AA }$. The EW of the edge-on galaxy is $\mathrm{ 2.5 \pm 0.7 \, \AA }$.

The \mgii\ lines of the face-on galaxies are much broader than those of the edge-on galaxies, because the central spectra of the face-on galaxies contain the down-the-barrel information on the galactic outflows along the line of sight.
The velocity difference of the two absorption lines thus indicates the typical velocity of the outflow ($v_{\mathrm{out}}$).
By measuring the \mgiia\ FWHM difference of the face-on and edge-on galaxies, we obtain an estimate of the typical outflow velocity of $\mathrm{161.5\,\pm 27.8 \, km/s}$.

We also provide another measurement of $v_{\mathrm{out}}$.
We perform a two-component Gaussian fit of the \mgiib\ absorption line, with a Gaussian component at zero velocity, and another Gaussian component with the velocity difference as a free parameter.
The first Gaussian represents the \mgii\ from the galaxy's interstellar medium (ISM), and the second one represents the galactic outflow.
The fit is shown by the red dotted line in Fig.~\ref{fig_spec_outflow2}.
The velocity difference of the two Gaussians is $\mathrm{199.4 \pm 19.6 \, km/s}$.
Although these two estimates of the outflow velocity are crude, they are in agreement within the error range.

\section{Physical properties of the galactic outflows.}

We have shown that the anisotropic \mgii\ emission extending up to $\approx$10~kpc is a common phenomenon for massive edge-on galaxies at $z \approx 1$, which demonstrates the prevalence of cool and metal-enriched galactic outflows.
The existence of outflows at high redshift is previously observed as the bimodal distribution of \mgii\ absorbers against bright background sources \cite{bouche12,zabl19,martin19,schroetter19,bordoloi11,kacprzak12,bordoloi14b,lan18,lundgren21}.
The bipolar pattern of outflows can also be statistically inferred by the down-the-barrel absorption.
The strengths and kinematics of the down-the-barrel absorbers are observed as a function of galaxy inclination angles \cite{chen10,kornei12,bordoloi14,rubin14}.
At similar redshift, galactic outflows are observed individually by emission lines in several cases \cite{rubin11,martin13,burchett21,shaban22,rupke19}.
Despite these individual findings, the occurrence frequency and bipolar shape of the galactic outflows, and their connection to the galaxy azimuthal angle are very unclear.
In our work, Fig.~\ref{fig_stack_img} provides the most direct evidence of the prevalence of cool and metal-enriched galactic outflows that form a bipolar geometry. 
In the previous sections, by imaging the galactic outflow in spatially-resolved spectroscopy, we have directly quantified its average morphology and kinematics.

Here we turn to provide an order-of-magnitude estimate of the \mgii\ density.
This estimate is based on the Sobolev approximation \cite{sobolev60, martin13,carr18,zabl21}.
The Sobolev approximation models the outflow with a radial velocity gradient.
Photons produced by star formation at frequency $\nu$ can only be resonant when they encounter a gas parcel with velocity $v(r_{s})=c(\nu-\nu_{0})/\nu_{0}$.
The radial velocity gradient thus ensures that the photons can only be resonant at the Sobolev radius $r_{s}$. 

In this work, we adopt the method of an identical case for an individual galaxy with \mgii\ emission \cite{martin13}.
Under the Sobolev approximation, the \mgii\ density can be estimated from the velocity gradient ${dv}/{dr}$ as:
\begin{equation}
\mathrm{
n_{Mg^{+}} \approx 6.9 \times 10^{-11} cm^{-3} \left| \frac{dv}{dr} \; \frac{kpc}{km\,s^{-1}} \right|_{r_{s}}
}
\end{equation}
Considering the outflow velocity $v_{\mathrm{out}} \approx 180 ~\mathrm{km/s}$, opening angle $\mathrm{ \theta \approx 70^{\circ } }$ and impact parameter $\mathrm{ b \approx 10 \, kpc }$, we finally get $\mathrm{n_{Mg^{+}} = 1.02 \times 10^{-9} \, cm^{-3}}$.

Given the ion density $\mathrm{n_{Mg^{+}}}$, the total gas density can be estimated by correcting for metallicity $\eta(\mathrm{Mg})$, ionization fraction $\chi(\mathrm{Mg^{+}})$ and dust depletion $d(\mathrm{Mg})$.
We assume solar metallicity, which is approximately the typical ISM metallicity for similar redshift and mass range \cite{zahid11}.
We chose a depletion typical of clouds in the Milky Way disk \cite{martin13}.
The ionization correction is a function of the ionization parameter, which depends on the ionizing photon luminosity Q.
\begin{equation}
\begin{aligned}
\mathrm{
n_{H} \approx 0.019 \, cm^{-3} 
\left( \frac{n_{Mg^{+}}}{10^{-9} \, cm^{-3}} \right)^{0.52} 
\left( \frac{10 \, kpc}{r_{s}} \right)^{0.96} 
\left( \frac{6.3\times 10^{-2}}{d(Mg)} \right)^{0.52} 
\left( \frac{3.8\times 10^{-5}}{\eta(Mg)} \right)^{0.52} 
\left( \frac{Q}{10^{53} \, s^{-1}} \right)^{0.48}
}
\end{aligned}
\end{equation}
Ref.~\cite{bacon23} provides measurements of the star formation rate based on spectral energy distribution fitting.
The median star formation rate of our sample is $\mathrm{\approx 5.1 \, M_\odot / yr}$.
We then estimate the typical Q of $\mathrm{\approx 4.7 \times 10^{53}\, s^{-1} }$ \cite{kennicutt98}.
Given our input value of $r_{s}$ and $\mathrm{n_{Mg^{+}}}$, we finally get our best estimate of the hydrogen density of $\mathrm{n_{H} = 0.03 \, cm^{-3}}$.

Finally, using the estimated H gas density and accounting for the solid angle of the outflow, we can estimate the average mass loss rate in the cool gas traced by \mgii\ \cite{martin13}.
\begin{equation}
\begin{aligned}
\mathrm{
\overset{.}{M}_{\rm out} = 30 \, M_{\odot} \, yr^{-1} \, \left(\frac{\Omega}{\pi} \right) \, \left(\frac{\mathit{f_c}}{1}\right) 
\left(\frac{r}{\rm 10 \, kpc}\right)^2 \, \left(\frac{v}{\rm 180 \, km\, s^{-1}}\right) \, \left(\frac{n_H}{\rm 0.03 \, cm^{-3}}\right)
}
\end{aligned}
\end{equation}
Here we assume the covering fraction $f_c\mathrm{ = 1}$.
We finally get $\mathrm{\overset{.}{M}_{\rm out}=36.3 \, M_{\odot}/yr}$.
This suggests that outflows typically remove $\approx$7 times more gas mass than is currently being converted into stars inside the galaxies. This roughly agrees with the predictions from simulations \cite{mitchell22}.

Note that here we only provide an order-of-magnitude estimate of $\mathrm{\overset{.}{M}_{\rm out}}$. 
As previously mentioned, the measurements of $\mathrm{\theta}$ and $\mathrm{b}$ are inevitably affected by the instrumental resolution.
For our stacked face-on galaxy sample, we make the assumption that the covering fraction $f_c$ of \mgii\ outflowing gas is uniformly 1. 
However, it is crucial to acknowledge that in real cases, $f_c$ may vary among different galaxies, thereby introducing non-linear contributions to the measurements.
Our estimate also depends on several parameters $\eta(\mathrm{Mg})$, $\chi(\mathrm{Mg^{+}})$, and $d(\mathrm{Mg})$, for which we do not know the exact values.
For example, we may expect the typical metallicity to be sub-solar.
If we assume a metallicity of $0.5 \, Z_{\odot}$, then $\mathrm{\overset{.}{M}_{\rm out}}$ would be boosted by a factor of 1.4.
Besides these parameters, $\mathrm{\overset{.}{M}_{\rm out}}$ depends mostly on $v_{\mathrm{out}}$, by a power index of 1.52.
To provide a possible range of $\mathrm{\overset{.}{M}_{\rm out}}$, 
we vary $v_{\mathrm{out}}$ from $\mathrm{170~\mathrm{km/s}}$ to $\mathrm{190~\mathrm{km/s}}$, and $\theta$ from $\mathrm{ 60^{\circ } }$ to $\mathrm{ 80^{\circ } }$. 
In that case $\mathrm{\overset{.}{M}_{\rm out}}$ varies from $\mathrm{23.9 \, M_{\odot}/yr}$ to $\mathrm{52.7 \, M_{\odot}/yr}$.


\end{methods}

\clearpage
\section*{Additional references}

\clearpage

\renewcommand{\figurename}{\textbf{Extended Data Fig.}}
\renewcommand{\tablename}{\textbf{Extended Data Table}}
\setcounter{figure}{0}
\newpage

\begin{figure}
\begin{center}
\includegraphics[width=0.95\textwidth]{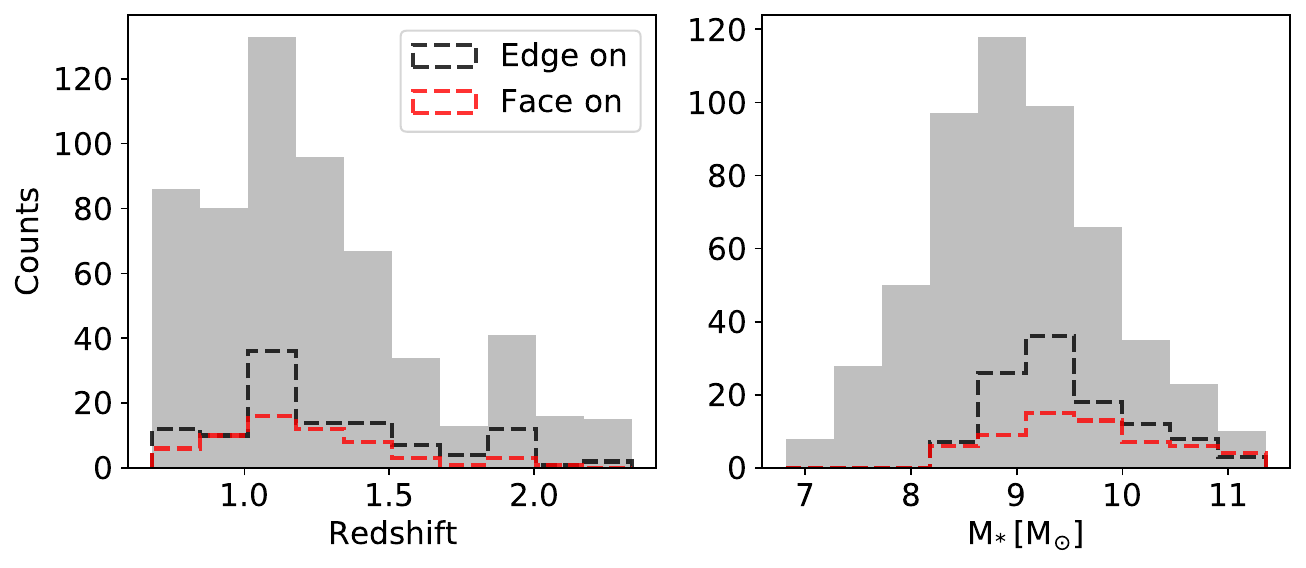}
\end{center}
\caption[]{\textbf{Distribution of the redshifts (left panel) and stellar masses (right panel) in the MUSE sample.}
The parent sample is shown by grey bars, and the edge-on and face-on subsamples are shown in black and red, respectively.  \label{fig_distribution}}
\end{figure}
\clearpage

\begin{figure}
\begin{center}
\includegraphics[width=0.95\textwidth]{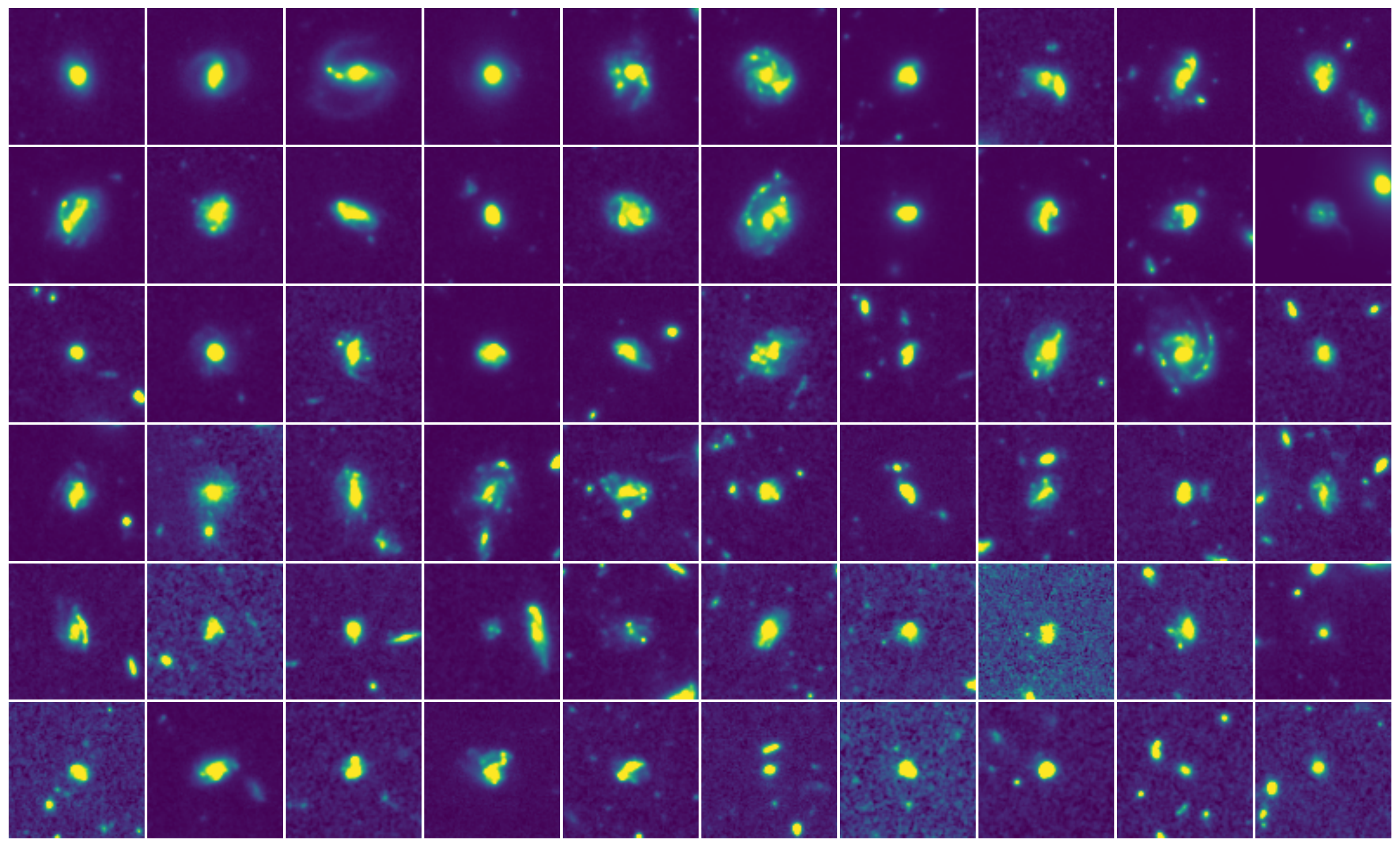}
\end{center}
\caption[]{\textbf{HST images of all the face-on galaxies.}  Each thumbnail has the same size as in Fig.~\ref{fig_stack_img}.
\label{fig_hst_fo} } 
\end{figure}
\clearpage

\begin{figure}
\begin{center}
\includegraphics[width=0.95\textwidth]{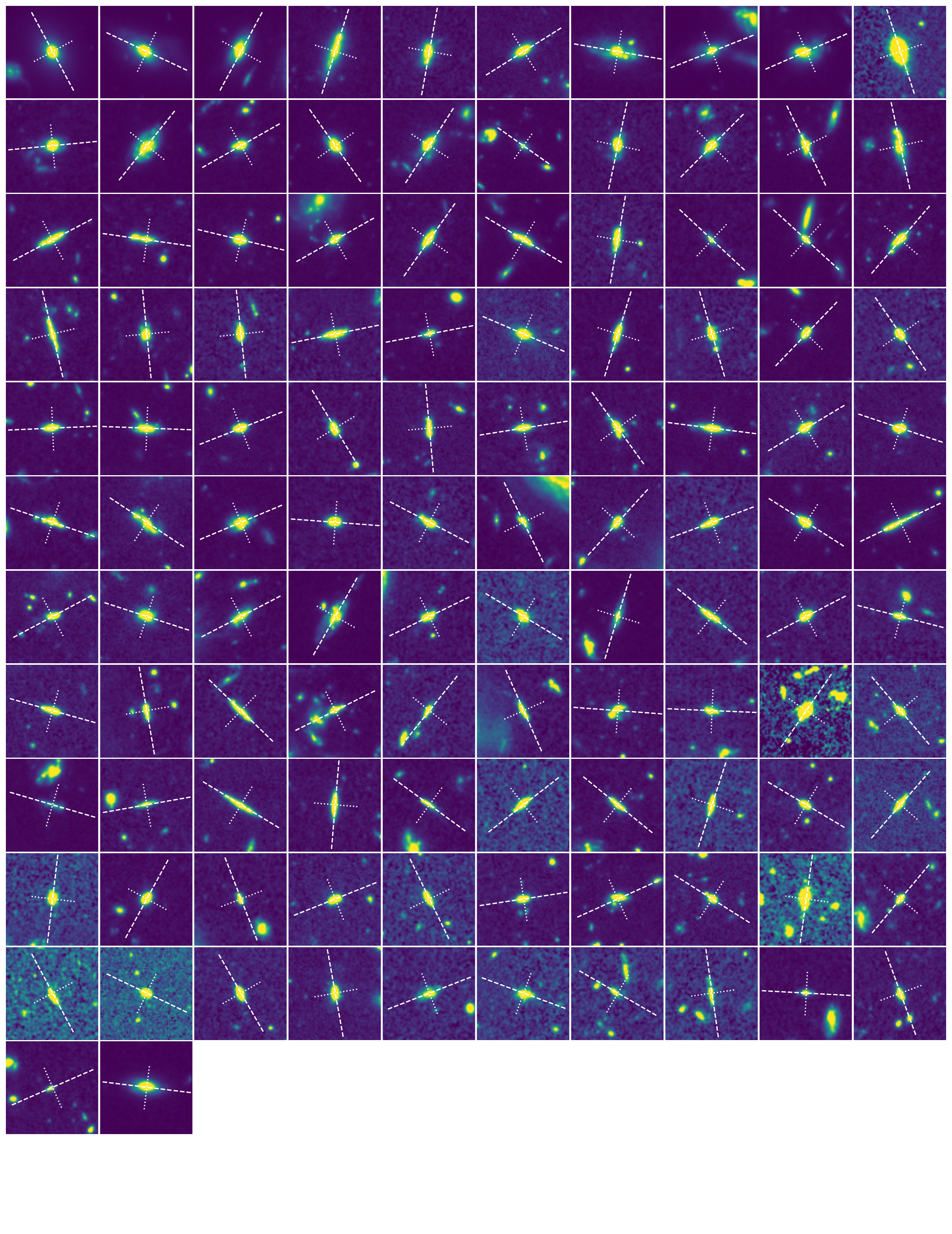}
\end{center}
\caption[]{\textbf{HST images of all the edge-on galaxies. }  Each thumbnail has the same size as in Fig.~\ref{fig_stack_img}. 
The dashed and dotted lines show the major and minor axes of the galaxies, respectively.
\label{fig_hst_eo} } 
\end{figure}
\clearpage

\begin{figure}
\begin{center}
\includegraphics[width=0.95\textwidth]{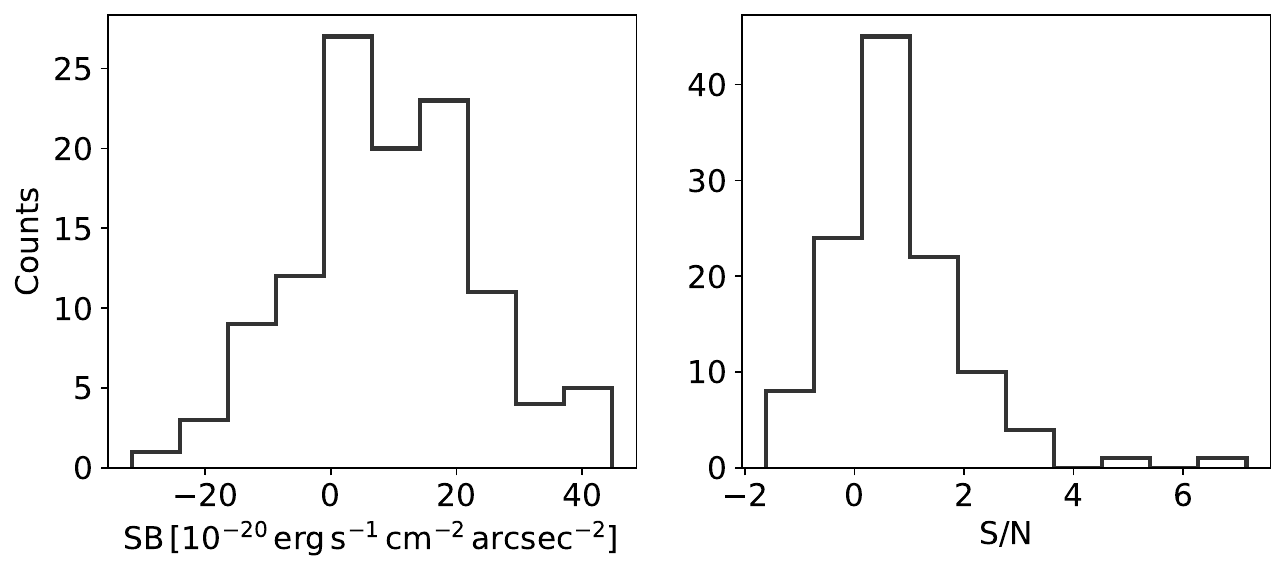}
\end{center}
\caption[]{\textbf{Distribution of the surface brightness (left panel) and S/N (right panel) of the \mgii\ outflow from each edge-on galaxy.}
The signals are extracted in 1''-diameter apertures above and below the edge-on galaxies, at a distance to galactic plane of 1''. 
The distribution of surface brightness signals skews towards positive values, despite the majority of the signals being of low S/N.
The negative S/N values in the right panel correspond to the negative signals in the left panel.
\label{fig_distribution_outflow}}
\end{figure}
\clearpage

\begin{figure}
\begin{center}
\includegraphics[width=0.7\textwidth]{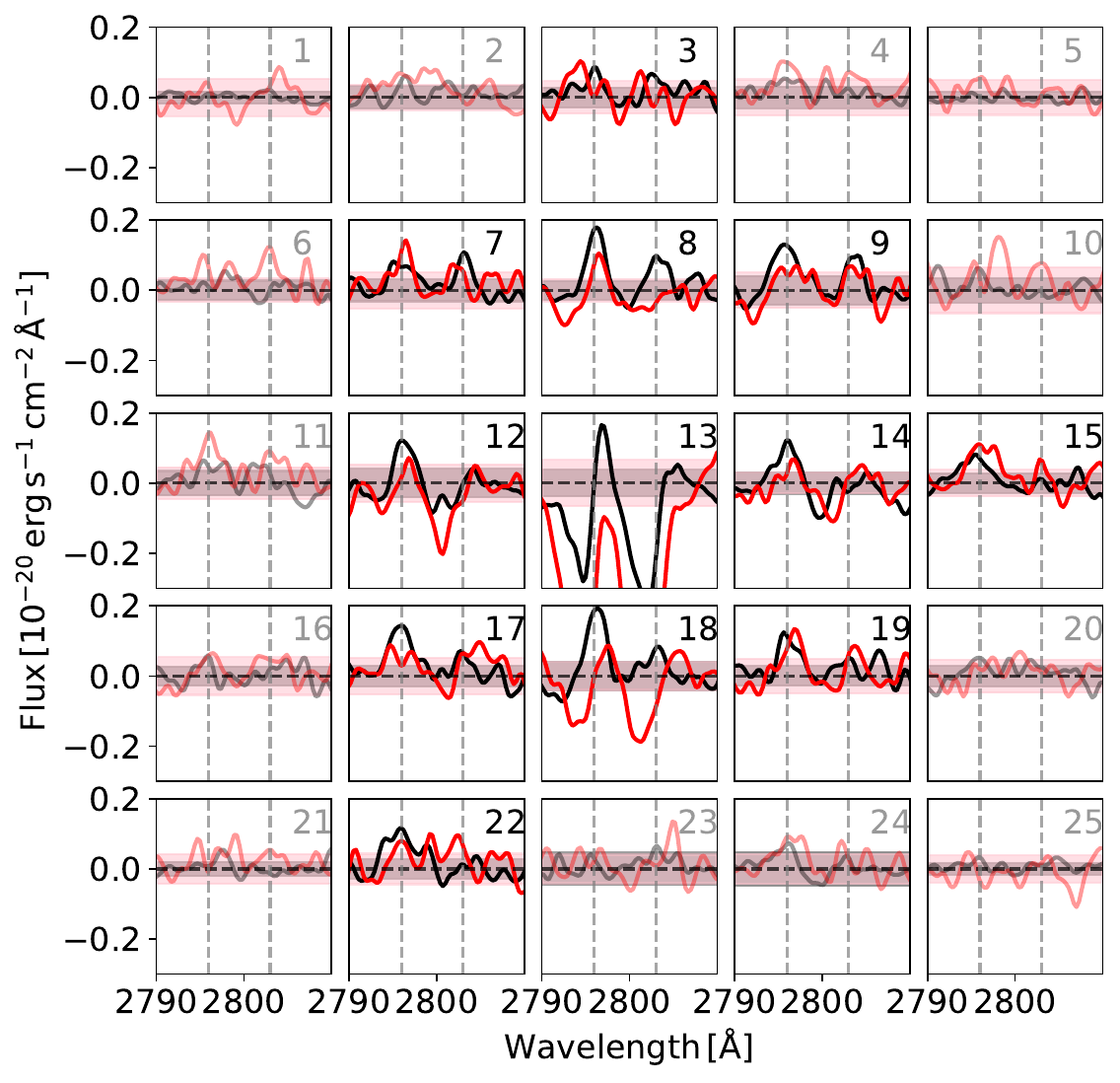}
\end{center}
\caption[]{\textbf{The spectra extracted from each grid cell at the corresponding position in Fig.~\ref{fig_stack_img}.}
The black and red spectra denote the edge-on and face-on galaxy samples, respectively. In each panel, the colored shading represents the 1$\sigma$ error range of the corresponding spectra. 
The panels for which the peak of the \mgii\ 2796 \AA\ line in the black spectra falls below the 2$\sigma$ threshold are marked with a lighter color.
\label{fig_stack_spec}} 
\end{figure}

\clearpage


\begin{figure}
\begin{center}
\includegraphics[width=0.7\textwidth]{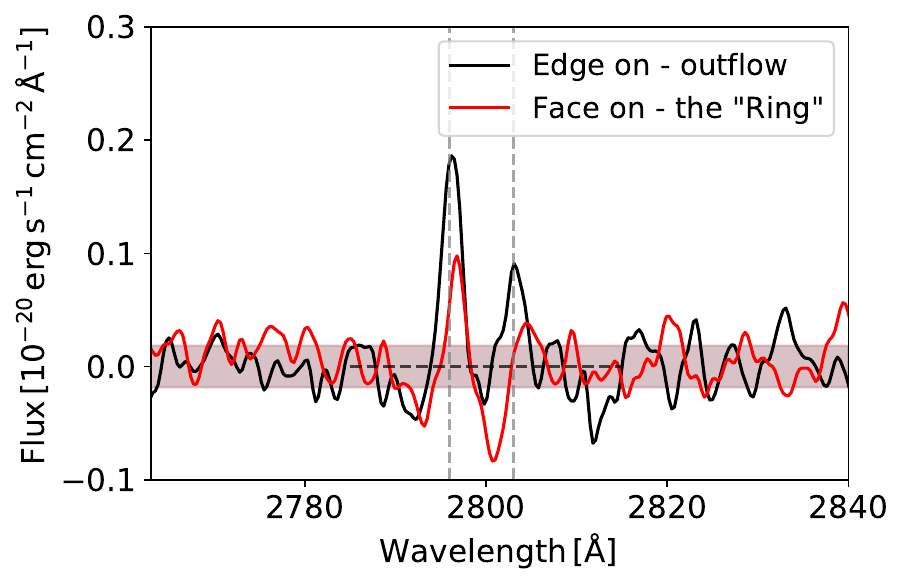}
\end{center}
\caption[]{\textbf{The continuum-subtracted spectrum of the ``ring" in face-on galaxies.}
The red line denotes the spectrum extracted from the ``ring" region of the stacked face-on galaxy sample.
For comparison, we also show the continuum-subtracted spectrum from the ``outflow" region of the edge-on galaxy.
The colored shadings represent the 1$\sigma$ error range of the corresponding spectra.
The two vertical dashed lines indicate the wavelength of \mgii\ doublets.
The horizontal shadow shows the noise level. 
The EWs of the \mgii\ 2796 \AA\ line for the red and black spectra are $\mathrm{ -21 \pm 15 \, \AA }$ and $\mathrm{ -56 \pm 22 \, \AA }$, respectively.

\label{fig_spec_outflow1}}
\end{figure}

\clearpage



\section*{References}

\vspace{5ex}

\begin{thebibliography}{10}
\expandafter\ifx\csname url\endcsname\relax
  \def\url#1{\texttt{#1}}\fi
\expandafter\ifx\csname urlprefix\endcsname\relax\def\urlprefix{URL }\fi
\providecommand{\bibinfo}[2]{#2}
\providecommand{\eprint}[2][]{\url{#2}}
\setlength{\itemsep}{0ex}


\bibitem{tumlinson17} Tumlinson, J., Peeples, M.~S., Werk, J.~K.\ 2017.\ The Circumgalactic Medium.\ Annual Review of Astronomy and Astrophysics 55, 389–432. doi:10.1146/annurev-astro-091916-055240

\bibitem{predehl20} Predehl, P. and 27 colleagues 2020.\ Detection of large-scale X-ray bubbles in the Milky Way halo.\ Nature 588, 227–231. doi:10.1038/s41586-020-2979-0

\bibitem{bland88} Bland, J., Tully, B.\ 1988.\ Large-scale bipolar wind in M82.\ Nature 334, 43–45. doi:10.1038/334043a0

\bibitem{muratov15} Muratov A.~L., Kere{\v{s}} D., Faucher-Gigu{\`e}re C.-A., Hopkins P.~F., Quataert E., Murray N., 2015, MNRAS, 454, 2691. doi:10.1093/mnras/stv2126

\bibitem{bacon10} Bacon, R. and 67 colleagues 2010.\ The MUSE second-generation VLT instrument.\ Ground-based and Airborne Instrumentation for Astronomy III 7735. doi:10.1117/12.856027

\bibitem{morrissey18} Morrissey, P. and 38 colleagues 2018.\ The Keck Cosmic Web Imager Integral Field Spectrograph.\ The Astrophysical Journal 864. doi:10.3847/1538-4357/aad597


\bibitem{wisotzki16} Wisotzki, L. and 24 colleagues 2016.\ Extended Lyman {\ensuremath{\alpha}} haloes around individual high-redshift galaxies revealed by MUSE.\ Astronomy and Astrophysics 587. doi:10.1051/0004-6361/201527384

\bibitem{wisotzki18} Wisotzki, L. and 24 colleagues 2018.\ Nearly all the sky is covered by Lyman-{\ensuremath{\alpha}} emission around high-redshift galaxies.\ Nature 562, 229–232. doi:10.1038/s41586-018-0564-6

\bibitem{leclercq17} Leclercq, F. and 19 colleagues 2017.\ The MUSE Hubble Ultra Deep Field Survey. VIII. Extended Lyman-{\ensuremath{\alpha}} haloes around high-z star-forming galaxies.\ Astronomy and Astrophysics 608. doi:10.1051/0004-6361/201731480

\bibitem{cai19} Cai, Z. and 8 colleagues 2019.\ Evolution of the Cool Gas in the Circumgalactic Medium of Massive Halos: A Keck Cosmic Web Imager Survey of Ly{\ensuremath{\alpha}} Emission around QSOs at z {\ensuremath{\approx}} 2.\ The Astrophysical Journal Supplement Series 245. doi:10.3847/1538-4365/ab4796

\bibitem{kusakabe22} Kusakabe, H. and 17 colleagues 2022.\ The MUSE eXtremely Deep Field: Individual detections of Ly{\ensuremath{\alpha}} haloes around rest-frame UV-selected galaxies at $\mathrm{z \approx 2.9-4.4}$.\ Astronomy and Astrophysics 660. doi:10.1051/0004-6361/202142302

\bibitem{bacon21} Bacon, R. and 31 colleagues 2021.\ The MUSE Extremely Deep Field: The cosmic web in emission at high redshift.\ Astronomy and Astrophysics 647. doi:10.1051/0004-6361/202039887

\bibitem{guo20} Guo, Y. and 10 colleagues 2020.\ Metal Enrichment in the Circumgalactic Medium and Ly{\ensuremath{\alpha}} Halos around Quasars at z {\ensuremath{\sim}} 3.\ The Astrophysical Journal 898. doi:10.3847/1538-4357/ab9b7f

\bibitem{johnson22} Johnson, S.~D. and 13 colleagues 2022.\ Directly tracing cool filamentary accretion over >100 kpc into the interstellar medium of a quasar host at z=1.\ arXiv e-prints.

\bibitem{herenz22} Herenz, E.~C. and 13 colleagues 2022.\ A $\sim$15 kpc Outflow Cone Piercing Through The Halo Of The Blue Compact Metal-poor Galaxy SBS0335-052.\ arXiv e-prints.

\bibitem{kacprzak13} Kacprzak, G.~G., Cooke, J., Churchill, C.~W., Ryan-Weber, E.~V., Nielsen, N.~M.\ 2013.\ The Smooth Mg II Gas Distribution through the Interstellar/Extra-planar/Halo Interface.\ The Astrophysical Journal 777. doi:10.1088/2041-8205/777/1/L11


\bibitem{rubin11} Rubin, K.~H.~R. and 6 colleagues 2011.\ Low-ionization Line Emission from a Starburst Galaxy: A New Probe of a Galactic-scale Outflow.\ The Astrophysical Journal 728. doi:10.1088/0004-637X/728/1/55

\bibitem{martin13} Martin, C.~L., Shapley, A.~E., Coil, A.~L., Kornei, K.~A., Murray, N., Pancoast, A.\ 2013.\ Scattered Emission from z \raisebox{-0.5ex}\textasciitilde 1 Galactic Outflows.\ The Astrophysical Journal 770. doi:10.1088/0004-637X/770/1/41

\bibitem{burchett21} Burchett, J.~N., Rubin, K.~H.~R., Prochaska, J.~X., Coil, A.~L., Vaught, R.~R., Hennawi, J.~F.\ 2021.\ Circumgalactic Mg II Emission from an Isotropic Starburst Galaxy Outflow Mapped by KCWI.\ The Astrophysical Journal 909. doi:10.3847/1538-4357/abd4e0

\bibitem{zabl21} Zabl, J. and 13 colleagues 2021.\ MusE GAs FLOw and Wind (MEGAFLOW) VIII. Discovery of a MgII emission halo probed by a quasar sightline.\ Monthly Notices of the Royal Astronomical Society 507, 4294–4315. doi:10.1093/mnras/stab2165

\bibitem{leclercq22} Leclercq, F. and 23 colleagues 2022.\ The MUSE eXtremely deep field: first panoramic view of an Mg II emitting intragroup medium.\ Astronomy and Astrophysics 663. doi:10.1051/0004-6361/202142179

\bibitem{bacon23} Bacon R., Brinchmann J., Conseil S., Maseda M., Nanayakkara T., Wendt M., Bacher R., et al., 2022, arXiv, arXiv:2211.08493

\bibitem{kornei12} Kornei, K.~A. and 7 colleagues 2012.\ The Properties and Prevalence of Galactic Outflows at z \raisebox{-0.5ex}\textasciitilde 1 in the Extended Groth Strip.\ The Astrophysical Journal 758. doi:10.1088/0004-637X/758/2/135

\bibitem{feltre18} Feltre, A. and 20 colleagues 2018.\ The MUSE Hubble Ultra Deep Field Survey. XII. Mg II emission and absorption in star-forming galaxies.\ Astronomy and Astrophysics 617. doi:10.1051/0004-6361/201833281

\bibitem{bouche12} Bouch{\'e}, N. and 6 colleagues 2012.\ Physical properties of galactic winds using background quasars.\ Monthly Notices of the Royal Astronomical Society 426, 801–815. doi:10.1111/j.1365-2966.2012.21114.x

\bibitem{zabl19} Zabl, J. and 12 colleagues 2019.\ MusE GAs FLOw and Wind (MEGAFLOW) II. A study of gas accretion around z {\ensuremath{\approx}} 1 star-forming galaxies with background quasars.\ Monthly Notices of the Royal Astronomical Society 485, 1961–1980. doi:10.1093/mnras/stz392

\bibitem{martin19} Martin, C.~L., Ho, S.~H., Kacprzak, G.~G., Churchill, C.~W.\ 2019.\ Kinematics of Circumgalactic Gas: Feeding Galaxies and Feedback.\ The Astrophysical Journal 878. doi:10.3847/1538-4357/ab18ac

\bibitem{schroetter19} Schroetter, I. and 12 colleagues 2019.\ MusE GAs FLOw and Wind (MEGAFLOW) - III. Galactic wind properties using background quasars.\ Monthly Notices of the Royal Astronomical Society 490, 4368–4381. doi:10.1093/mnras/stz2822

\bibitem{muratov17} Muratov, A.~L. and 10 colleagues 2017.\ Metal flows of the circumgalactic medium, and the metal budget in galactic haloes.\ Monthly Notices of the Royal Astronomical Society 468, 4170–4188. doi:10.1093/mnras/stx667

\bibitem{peroux20} P{\'e}roux, C. and 6 colleagues 2020.\ Predictions for the angular dependence of gas mass flow rate and metallicity in the circumgalactic medium.\ Monthly Notices of the Royal Astronomical Society 499, 2462–2473. doi:10.1093/mnras/staa2888



\end{thebibliography}

\vspace{5ex}


\begin{thebibliography}{10}
\expandafter\ifx\csname url\endcsname\relax
  \def\url#1{\texttt{#1}}\fi
\expandafter\ifx\csname urlprefix\endcsname\relax\def\urlprefix{URL }\fi
\providecommand{\bibinfo}[2]{#2}
\providecommand{\eprint}[2][]{\url{#2}}
\setlength{\itemsep}{0ex}
\setcounter{enumiv}{30}

\bibitem{claeyssens19} Claeyssens, A. and 15 colleagues 2019.\ Spectral variations of Lyman {\ensuremath{\alpha}} emission within strongly lensed sources observed with MUSE.\ Monthly Notices of the Royal Astronomical Society 489, 5022–5029. doi:10.1093/mnras/stz2492

\bibitem{leclercq20} Leclercq, F. and 20 colleagues 2020.\ The MUSE Hubble Ultra Deep Field Survey. XIII. Spatially resolved spectral properties of Lyman {\ensuremath{\alpha}} haloes around star-forming galaxies at z $>$ 3.\ Astronomy and Astrophysics 635. doi:10.1051/0004-6361/201937339

\bibitem{erb22} Erb, D.~K. and 7 colleagues 2022.\ The Circumgalactic Medium of Extreme Emission Line Galaxies at z \raisebox{-0.5ex}\textasciitilde 2: Resolved Spectroscopy and Radiative Transfer Modeling of Spatially Extended Lyman-alpha Emission in the KBSS-KCWI Survey.\ arXiv e-prints.

\bibitem{bacon17} Bacon, R. and 26 colleagues 2017.\ The MUSE Hubble Ultra Deep Field Survey. I. Survey description, data reduction, and source detection.\ Astronomy and Astrophysics 608. doi:10.1051/0004-6361/201730833


\bibitem{bouche21} Bouch{\'e}, N.~F. and 16 colleagues 2021.\ The MUSE Hubble Ultra Deep Field Survey. XVI. The angular momentum of low-mass star-forming galaxies: A cautionary tale and insights from TNG50.\ Astronomy and Astrophysics 654. doi:10.1051/0004-6361/202040225

\bibitem{rubin14} Rubin, K.~H.~R., Prochaska, J.~X., Koo, D.~C., Phillips, A.~C., Martin, C.~L., Winstrom, L.~O.\ 2014.\ Evidence for Ubiquitous Collimated Galactic-scale Outflows along the Star-forming Sequence at z \raisebox{-0.5ex}\textasciitilde 0.5.\ The Astrophysical Journal 794. doi:10.1088/0004-637X/794/2/156

\bibitem{walter02} Walter, F., Weiss, A., Scoville, N.\ 2002.\ Molecular Gas in M82: Resolving the Outflow and Streamers.\ The Astrophysical Journal 580, L21–L25. doi:10.1086/345287

\bibitem{chisholm20} Chisholm, J., Prochaska, J.~X., Schaerer, D., Gazagnes, S., Henry, A.\ 2020.\ Optically thin spatially resolved Mg II emission maps the escape of ionizing photons.\ Monthly Notices of the Royal Astronomical Society 498, 2554–2574. doi:10.1093/mnras/staa2470

\bibitem{katz22} Katz, H. and 9 colleagues 2022.\ Mg II in the JWST era: a probe of Lyman continuum escape?.\ Monthly Notices of the Royal Astronomical Society 515, 4265–4286. doi:10.1093/mnras/stac1437

\bibitem{bordoloi11} Bordoloi, R. and 52 colleagues 2011.\ The Radial and Azimuthal Profiles of Mg II Absorption around 0.5 < z < 0.9 zCOSMOS Galaxies of Different Colors, Masses, and Environments.\ The Astrophysical Journal 743. doi:10.1088/0004-637X/743/1/10

\bibitem{kacprzak12} Kacprzak, G.~G., Churchill, C.~W., Nielsen, N.~M.\ 2012.\ Tracing Outflows and Accretion: A Bimodal Azimuthal Dependence of Mg II Absorption.\ The Astrophysical Journal 760. doi:10.1088/2041-8205/760/1/L7

\bibitem{bordoloi14b} Bordoloi, R., Lilly, S.~J., Kacprzak, G.~G., Churchill, C.~W.\ 2014.\ Modeling the Distribution of Mg II Absorbers around Galaxies Using Background Galaxies and Quasars.\ The Astrophysical Journal 784. doi:10.1088/0004-637X/784/2/108

\bibitem{lan18} Lan, T.-W., Mo, H.\ 2018.\ The Circumgalactic Medium of eBOSS Emission Line Galaxies: Signatures of Galactic Outflows in Gas Distribution and Kinematics.\ The Astrophysical Journal 866. doi:10.3847/1538-4357/aadc08

\bibitem{lundgren21} Lundgren, B.~F. and 12 colleagues 2021.\ The Geometry of Cold, Metal-enriched Gas around Galaxies at z {\ensuremath{\sim}} 1.2.\ The Astrophysical Journal 913. doi:10.3847/1538-4357/abef6a

\bibitem{chen10} Chen, Y.-M. and 6 colleagues 2010.\ Absorption-line Probes of the Prevalence and Properties of Outflows in Present-day Star-forming Galaxies.\ The Astronomical Journal 140, 445–461. doi:10.1088/0004-6256/140/2/445

\bibitem{bordoloi14} Bordoloi, R. and 46 colleagues 2014.\ The Dependence of Galactic Outflows on the Properties and Orientation of zCOSMOS Galaxies at z \raisebox{-0.5ex}\textasciitilde 1.\ The Astrophysical Journal 794. doi:10.1088/0004-637X/794/2/130

\bibitem{shaban22} Shaban, A. and 12 colleagues 2022.\ A 30 kpc Spatially Extended Clumpy and Asymmetric Galactic Outflow at z   1.7.\ The Astrophysical Journal 936. doi:10.3847/1538-4357/ac7c65

\bibitem{rupke19} Rupke, D.~S.~N. and 11 colleagues 2019.\ A 100-kiloparsec wind feeding the circumgalactic medium of a massive compact galaxy.\ Nature 574, 643–646. doi:10.1038/s41586-019-1686-1

\bibitem{sobolev60} Sobolev, V.~V.\ 1960.\ Moving Envelopes of Stars.\ Moving Envelopes of Stars, by V. V. Sobolev, Translated by Sergei Gaposchkin, Copyright: 1960, eBook: 2013, Reprint: 2014. Cambridge: Harvard University Press. OCLC: 1013938845. ISBN: 9780674864634, eISBN: 9780674864658.. doi:10.4159/harvard.9780674864658

\bibitem{carr18} Carr, C., Scarlata, C., Panagia, N., Henry, A.\ 2018.\ A Semi-analytical Line Transfer (SALT) Model. II: The Effects of a Bi-conical Geometry.\ The Astrophysical Journal 860. doi:10.3847/1538-4357/aac48e

\bibitem{zahid11} Zahid, H.~J., Kewley, L.~J., Bresolin, F.\ 2011.\ The Mass-Metallicity and Luminosity-Metallicity Relations from DEEP2 at z \raisebox{-0.5ex}\textasciitilde 0.8.\ The Astrophysical Journal 730. doi:10.1088/0004-637X/730/2/137


\bibitem{kennicutt98} Kennicutt, R.~C.\ 1998.\ Star Formation in Galaxies Along the Hubble Sequence.\ Annual Review of Astronomy and Astrophysics 36, 189–232. doi:10.1146/annurev.astro.36.1.189

\bibitem{mitchell22} Mitchell, P.~D., Schaye, J.\ 2022.\ How gas flows shape the stellar-halo mass relation in the EAGLE simulation.\ Monthly Notices of the Royal Astronomical Society 511, 2948–2967. doi:10.1093/mnras/stab3339



\end{thebibliography}
\end{document}